\documentclass[aps,pra,twocolumn,preprintnumbers,amsmath,amssymb,superscriptaddress,floatfix]{revtex4-1}
\usepackage[utf8]{inputenc}
\usepackage{graphicx}% Include figure files
\usepackage{dcolumn}% Align table columns on decimal point
\usepackage{bm}% bold math
\usepackage[english]{babel}
\usepackage{amsmath}
\usepackage{amssymb}
\usepackage{amstext}
\usepackage{braket}
\usepackage{bbold}
\global\long\def\proj#1{\mbox{\ensuremath{|#1\rangle\!\langle#1|}}}
\begin{document}
%opening
\title{Simulation of non-Abelian gauge theories with optical lattices}
\author{L. Tagliacozzo}
\email{luca.tagliacozzo@icfo.es} \affiliation{ICFO - The Institute
of Photonic Sciences, Av. C.F. Gauss 3, E-08860 Castelldefels
(Barcelona), Spain}
\author{A. Celi}
\email{alessio.celi@icfo.es} \affiliation{ICFO - The Institute of
Photonic Sciences, Av. C.F. Gauss 3, E-08860 Castelldefels
(Barcelona), Spain}
\author{P. Orland}
\affiliation{Baruch College and the Graduate School and University Center, CUNY, New York, NY 10010, USA}
\author{M. Mitchell}
\affiliation{ICFO - The Institute of Photonic Sciences, Av. C.F.
Gauss 3, E-08860 Castelldefels (Barcelona), Spain}
\author{M. Lewenstein}
\affiliation{ICFO - The Institute of Photonic Sciences, Av. C.F.
Gauss 3, E-08860 Castelldefels (Barcelona), Spain}
\affiliation{ICREA-Instituci\'{o} Catalana de Recerca i Estudis
Avancats, 08010 Barcelona, Spain}

\maketitle

{\bf Many phenomena occurring in strongly correlated quantum
systems still await  conclusive explanations. The absence
of isolated free quarks  in nature is an example. It is
attributed to {\em quark confinement}, whose origin is not yet understood.
The phase diagram
for nuclear matter at general temperatures and densities, studied in heavy-ion collisions, is not settled.
Finally, we have no definitive theory of high-temperature superconductivity. Though
we have theories that could underlie such physics, we 
lack the tools to determine the experimental consequences of
these theories. {\em Quantum simulators} may provide such tools. Here we
show how to engineer  quantum simulators of 
non-Abelian lattice gauge theories. The systems we consider have
several applications: they can be used 
to mimic quark confinement or to
study dimer and valence-bond states (which may be relevant for
high-temperature superconductors)}.

Gauge theories (GT) \cite{oraifeartaigh_gauge_2000} provide the
basis of modern physics. In the ``Standard Model of Particle
Physics" \cite{gaillard_standard_1999} GT
describe three of the four fundamental interactions (namely
``electromagnetic'', ``weak'', and ``strong'' interactions. The
last  is described by the GT known as Quantum Chromo-Dynamics (QCD)
\cite{smilga_lectures_2001}). Gauge symmetry plays also a
role in General Relativity. At the same time, GT
are present in many effective models of condensed matter, {\em e.g.}
antiferromagnets \cite{balents_spin_2010} and high-temperature
superconductors \cite{lee_doping_2006,mann_high-temperature_2011}.
Recently, the study of phase diagrams of various GT  has gained new attention,
because of the discovery of topological order. Due to their stability against perturbations, topologically-ordered
phases may help to design quantum computers
\cite{kitaev_fault-tolerant_1997,trebst_breakdown_2006,tupitsyn_topological_2008, tagliacozzo_entanglement_2011,dusuel_robustness_2011}.

Despite the enormous importance of GT, they defy solution. Wilson's 
formulation
\cite{wilson_confinement_1974} of lattice gauge theories
(LGT), where continuous space-time is replaced by a discrete set
of points, provided the first numerical tool to study the
strong coupling regime. Monte-Carlo
(MC) simulations of LGT is the main 
tool to compare  aspects of QCD at strong-coupling  with experiments
\cite{bazavov_nonperturbative_2010}. What is hard or
impossible to compute with MC remains out of reach. For
example, the mechanism of charge confinement
\cite{polyakov_compact_1975}, invented to explain the absence of
isolated quarks \cite{kim_search_2007}, is still debated
four decades since first proposed. Furthermore, MC simulations cannot 
yet provide definite predictions for hot and dense nuclear matter \cite{satz_sps_2004,gupta_scale_2011}, probed
by heavy nuclei collisions at CERN  and RHIC \cite{alice_collaboration_j/_2012,star_collaboration_directed_2012}. GT are also
invoked in explanations of spin-liquid phases of
antiferromagnets \cite{balents_spin_2010} and high-temperature
superconductivity \cite{anderson_resonating_1987}.

Recent progress in the experimental control of quantum systems makes 
possible to engineer systems that perfectly mimic theoretical
models. This is the idea of {\em quantum simulators}
\cite{lewenstein_ultracold_2012,hauke_can_2012,bloch_quantum_2012,blatt_quantum_2012,
aspuru-guzik_photonic_2012,houck_-chip_2012}, whose ultimate goal
is to simulate GT, {\rm e.g.} QCD,and provide access to their
phase diagrams at finite temperature and density. A more modest
goal is to emulate QCD, {\em i.e.} to find a model sharing its interesting properties, whose
realization may be simpler than the full theory.
The first steps of this emulation program were to
describe quantum simulations of Abelian LGT
\cite{zohar_confinement_2011,zohar_simulating_2012,tagliacozzo_optical_2012,banerjee_atomic_2012,zohar_simulating_2012_b,zohar_topological_2012}.
The presence  of many-body interactions, beyond nearest
neighbors, has been the main technical obstacle. This obstacle has been addressed in  
\cite{tagliacozzo_optical_2012}, by using
mesoscopic Rydberg gates \cite{Muller09}.

Here we show how to simulate {\it non-Abelian gauge magnets} (GM)
or {\it link models}, introduced in  
\cite{horn_finite_1981,orland_lattice_1990,Chandrasekharan:1996ih} using Rydberg atoms
\cite{Weimer10,schaus_observation_2012}.
The models have both strong- and weak-coupling regimes. We 
discuss here the origin of charge confinement in
both regimes, stressing the different physical origin in each. We propose 
how to identify the flux-tubes connecting
static external charges in each of these regimes, and provide the experimental protocol to
observe these flux tubes. We conclude  by discussing  a qualitative technique,
based on energy landscapes around static charges, to
identify in a generic LGT whether  chromo-electric strings, {\em i.e.}
charge confinement, is present.

\section{The model}
Here we want to analyze a specific non-Abelian gauge theory that can be simulated with ultra-cold atoms. 
Gauge theories were originally introduced in the context of relativistic field theories as generalization of quantum electrodynamics (QED), the theory of photons and electrons, and hence formulated 
through a Lagrangian density that does not distinguish between space and time.
In order study them with   quantum simulators, we need  their  Hamiltonian formulation \cite{kogut_hamiltonian_1975,creutz_gauge_1977} on the lattice.
%In the experiment, indeed  the lattice is given by  an optical lattice loaded  with neutral atoms and the time is the  real time during which the experiment will run.
There, the constituents representing the gauge  bosons (generalization of photons)   live on the links of the lattice and those
representing the charged matter (generalization of electrons) live on the sites.

The Hamiltonian of a LGT is manifestly invariant under a group of local transformations. These transformations encode the generalization of the Gauss law, that physically enforces the charge conservation. 
The choice of the symmetry group determines if we are dealing with an Abelian (for example QED) or non-Abelian LGT. One of the  simplest non-Abelian LGT is the $SU(2)$ LGT.
The specific form of the  Hamiltonian and the  Hilbert space of the constituents determines which  $SU(2)$ LGT one considers . 
The standard $SU(2)$ LGT, called Yang-Mills theory (YM), involves an infinite dimensional Hilbert space for the gauge bosons and an Hamiltonian  obtained  directly from the original  Wilson action \cite{wilson_confinement_1974}.

In a quantum simulation we  encode the states of the constituents in  hyperfine levels of the atoms. For this reason we want to study first the simplest $SU(2)$ LGT having as  small as possible local Hilbert space.  This leads to the family of the  $SU(2)$ GM. Although their physical properties are quite different from the one of the YM theory, the $SU(2)$ GM   we analyze here  shares with YM the phenomenon of confinement of charges.
% In order to address this phenomenon we can restrict our attention to gauge boson sector  of the  $SU(2)$  theory in presence of a static background of external charges.
% A linear increase of the ground state energy of the gauge system with the separation  of two external charges, will indeed signal their  confinement.

In the  $SU(2)$ GM, gauge bosons have four states, so that their Hilbert space is isomorphic ${\cal C}^2\otimes {\cal C}^2$.
In order to impose the Gauss law we first need to define the operators that implement the rotation of an arbitrary group element. An element of $SU(2)$ can be written as $\exp  [i \vec{\alpha} \cdot \vec{\sigma} ]$ where $\vec{\alpha} =(\alpha_1, \alpha_2, \alpha_3)$ is real three-component vector 
and $\vec\sigma=(\sigma_1,\sigma_2,\sigma_3)$ are the Pauli matrices.

The rotation of a link state  through a group element is obtained with the two operators,  
\begin{eqnarray}
 \Xi(\vec{\alpha}) = \proj{0} \otimes \exp (i \vec{\alpha} \cdot \vec{\sigma}) + \proj{1} \otimes I,  \label{eq:S1} \\
 \tilde{\Xi}(\vec{\alpha}) =  \proj{0}
 \otimes  I +\proj{1} \otimes \exp (i \vec{\alpha} \cdot \vec{\sigma})  \label{eq:S2},
\end{eqnarray}

\begin{figure}
\includegraphics[width=7cm]{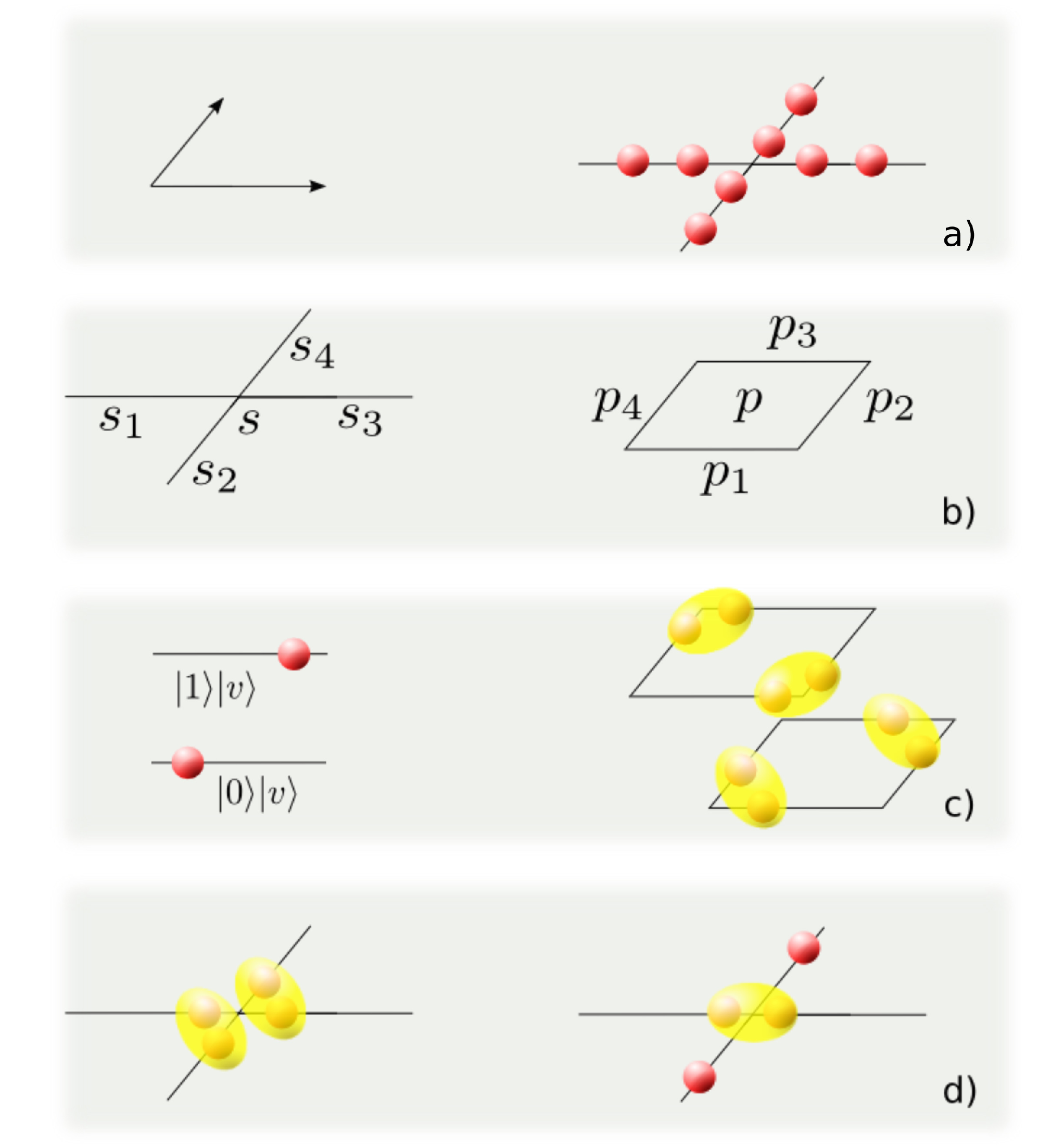}
\caption{\label{fig:ingredients}\emph{The basics of 
non-Abelian GM.}  a) {\bf Hilbert space}, we assign to each link of  an oriented lattice (left), the
Hilbert space of two qubits, each of
them represented by a sphere (right). b) {\bf Notation}, a link may be labeled in either of 
two equivalent ways: by adding a subscript referring to the
adjacent site (left panel), or to the adjacent plaquette (right panel). We number these
labels counterclockwise. c) Left-panel, {\bf Physical interpretation},  representation of  ``position'' and `spin'' degrees of freedom, a red sphere on the left of a link represents the state  $\ket{0} \ket{v}$,
while a red sphere on the right of a link represents the state $\ket{1} \ket{v}$.
c) Right-panel, {\bf Gauge invariant space} for a plaquette, gauge  invariance
forces qubits adjacent to a site to form singlets (yellow ovals
around them).  d) {\bf Gauge invariant states}, example of gauge invariant state for a generic lattice.}
 \end{figure}

In this way we  can define a local  transformation in absence of external charges as
\begin{equation}
G_s(\vec{\alpha})= \tilde{\Xi}(\vec{\alpha})_{s_1} \otimes \tilde{\Xi}(\vec{\alpha})_{s_2}
 \otimes  {\Xi}(\vec{\alpha})_{s_3} \otimes {\Xi}(\vec{\alpha})_{s_4}, \label{symm}
\end{equation}
on the ``crosses'', which consist of the four spin states adjacent to given site $s$ (labeled $s_1 ,\dots, s_4$, 
see Fig. \ref{fig:ingredients}~b)). The gauge-invariance constraint, playing the role of the QED Gauss law, selects those states with conserved charge,   
\begin{equation}
\set{\ket{\psi}}: G_s(\vec{\alpha})\ket{\psi} = \ket{\psi}, \forall  s, \vec{\alpha}  \label{eq:g-inv},
\end{equation}
that are the only physical states of the LGT.
It is sufficient to impose this condition for $\vec \alpha$ equal to $\hat i=(1,0,0)$, $\hat j=(0,1,0)$, and $\hat k=(0,0,1)$.

We can  derive the physical consequences of (\ref{eq:g-inv}). We interpret the local Hilbert space as  $\cal{P} \otimes \cal{S}$,
describing one qubit (the right factor, ${\cal S}$) moving between the two
ends of the link (the left factor, ${\cal P}$).
We identify  the basis of ${\cal P}$, $\ket{0}$ and   $\ket{1}$,  with
the left-end  (lower-end) or the  right-end (upper-end)  of a link in the $x$ ($y$) direction.
In this way,  given a generic state  $\ket{v}$  in $\cal{S}$, we represent  $\ket{0}\ket{v}$  ($\ket{1}\ket{v}$) as a solid dot on the left (right) part
of the link, cf. Fig \ref{fig:ingredients}~c). The operator
$\Xi(\vec\alpha)$ acts  on those vectors in the left (down)
two-dimensional  subspace
$\ket{0}\set{\ket{v}}$ of the $x$ ($y$) oriented
link (rotating them by $\exp{(i \vec\alpha\cdot
\vec\sigma )}$). The operator $\tilde{\Xi}(\vec\alpha)$ acts similarly on the other subspace of that link.
Hence, the physical-state condition (\ref{eq:g-inv}) forces the total
spin of the qubits adjacent to the site $s$ to be zero, {\em i.e.} to consists
 of singlet among pairs of those, $S_{ij}\equiv \frac
1{\sqrt 2} (\uparrow\downarrow - \downarrow\uparrow)$, cf. Fig.
\ref{fig:ingredients}~d).

Charges are encoded  by additional  qubits on the sites of the lattice. Different spin representations $\vec S=1/2,1,3/2,2$ require the addition of a different  amount of qubits. The presence of charge at the site $\tilde s$ implies that the gauge transformations at  $\tilde s$  induce a rotation of the state, 
\begin{equation}
G_{\tilde s}(\vec{\alpha}) \ket{\psi} = \exp {(i\vec{\alpha} \cdot \vec{S}) }\ket{\psi}, \forall \vec{\alpha}, \label{eq:ext-charge}
\end{equation}
Here we focus on the case of spin-1/2 charges, {\em i.e.} 
${\vec S}\equiv \vec\sigma$.
In this case the charge is encoded by a qubit located at $\tilde{s}$ and (\ref{eq:ext-charge}) can be expressed as
\begin{equation}
G^{ext}_{\tilde s}(\vec{\alpha}) \ket{\psi} = \ket{\psi}, \forall \vec{\alpha} \label{eq:g-inv-ext} 
\end{equation}
with $G^{ext}_{\tilde s}(\vec{\alpha}) =G_{\tilde s}(\vec{\alpha}) \otimes \exp {(i\vec{\alpha}\cdot \vec{\sigma}_{\tilde{s}})}$.

 We now turn to the form of the Hamiltonian $H$. We built $H$ as a sum of plaquette, $H_p$, and link, $H_l$, operators as for ordinary LGT ($p$ and $l$ label the plaquettes and the links of the lattice). 
 The former corresponds to the magnetic term ($B^2$ in electrodynamics) while the latter is the analog of the electric term ($E^2$).
 A convenient set of operators to write $H_p$ and $H_l$ is (see  \cite{orland_lattice_1990})
\begin{equation}
\Gamma^0 = \sigma_1 \otimes I,\  \Gamma^j = \sigma_2 \otimes \sigma_j,\;j=1,2,3, \
\Gamma^5 = \sigma_3 \otimes I.
\label{eq:D}
\end{equation}
Gauge invariance, i.e. $[G_s, H] =0, \forall s$, fixes $H_l$ and $H_p$.
It immediately implies $[\Xi(\vec{\alpha}) ,H_l] =0$, thus, $H_l\propto\Gamma^5$ (plus trivial identity term).
$H_p$ can be written as ${\rm tr}_V (U_{p_1} \otimes U_{p_2} \otimes {U_{p_3}}^{\dagger} \otimes {U_{p_4}}^{\dagger} )$,
$p_1 ,\dots\;, p_4$ being the links around the plaquette $p$, ordered as in Fig. \ref{fig:ingredients}~b).
At each link, the $U$ acts on $({\cal P}\otimes{\cal S})\otimes V$, where $V$ is an auxiliary bookkeeping  spin-1/2 space on which the trace is performed.
$V$ is not physically  implemented in the simulator and is introduced only to write covariant expression easily. 
The gauge invariance requirement for $U$ reads 
\begin{equation}
 [\Xi(\vec{\alpha}) ,U] = \exp (i \vec{\alpha} \cdot \vec{\tau} ) U ,\; [U,  \tilde{\Xi}(\vec{\alpha})] =  U \exp (-i \vec{\alpha} \cdot \vec{\tau} ),\label{eq:gtU}
\end{equation}
where $\tau^j$, $j=1,2,3$ is a Pauli matrix on $V$. Again, (\ref{eq:gtU}) has the same form as in standard LGT.
A solution of (\ref{eq:gtU}) is $U=\Gamma^0 \otimes 1 -i \sum_{j=1}^3 \Gamma^j \otimes \tau^j$, that is the one we consider here. Thus,
\begin{equation}
\frac H{\Delta} = 
\sum_l \Gamma^5_l + 
\frac{1}{g} 
\sum_p {\rm tr}_V (U_{p_1} \otimes U_{p_2} \otimes {U_{p_3}}^{\dagger} \otimes {U_{p_4}}^{\dagger} ) +H.c., \label{eq:ham}
\end{equation} where $\Delta$ is an energy scale. 
The coupling constant $g$ determines if the system is in the \emph{weak-} or in the \emph{strong-coupling} regime ($g\to 0$ and $g\to \infty$, respectively).

\section{The confinement phase} %LUCA, I have changed the heading for this section.

The phases of LGT are commonly characterized through the force induced by gauge bosons on the charges.
Here we are interested in studying the confinement phase, that is the phase where the attractive force between two charges, does not depend on their distance.
In this case, the ground state energy of the system in presence of  two charges, increases linearly with their separation.

\begin{figure}
\includegraphics[width=7cm]{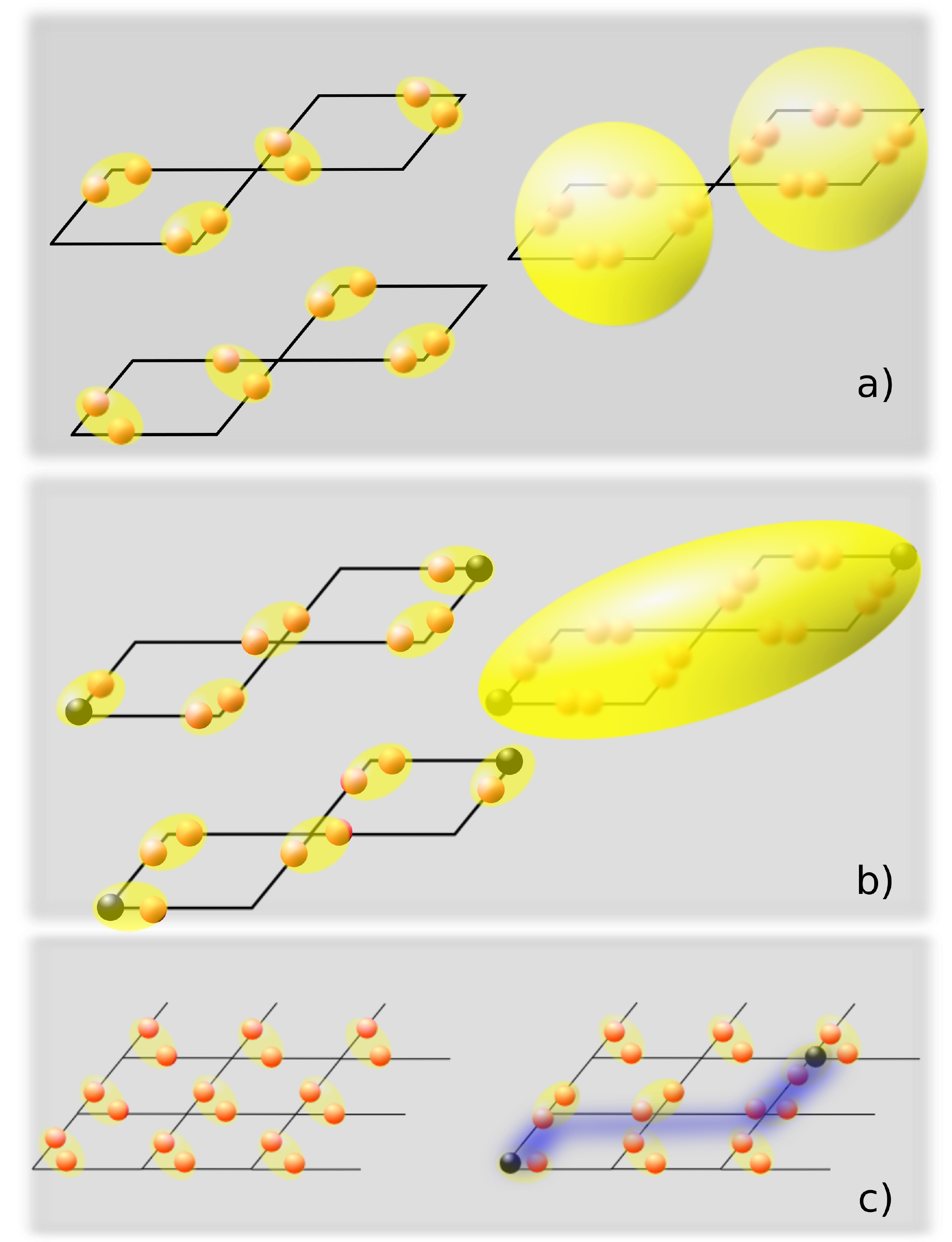}
\caption{\label{fig:wc-conf}\emph{Confinement of charges}.
a), b) {\bf Weak coupling}. a) In absence of external charges,
dimers resonate on configurations allowed by gauge invariance. b)  When two static charges (black spheres) are inserted,
plaquettes are entangled by strings of singlets.  In this way, a ``macromolecule" or ``polymer", as large as the separation of the charges, is formed (yellow 
oval). The energy of such a state 
rises linearly with the inter-charge separation, thereby confining charges.
c)  {\bf Strong-coupling}. The
ground state of (\ref{eq:ham}) is the ordered configuration where all position qubits are in the state $\ket{0}$ and spin qubits are forced by (\ref{eq:g-inv}) to form singlets (left panel). When two external charges are inserted (black spheres), the
singlets have to rearrange (right panel). Some of the position qubits are flipped to $\ket{1}$  with an associated energy cost. Pictorially the corresponding spin qubits shift right-up on
the $x$-$y$ links. A string can be
identified by position qubits in state  $\ket{1}$. The energy cost of a string is
proportional to its length, hence, the charges are confined.}
\end{figure}

At \emph{weak-coupling}, $H$ in (\ref{eq:ham}) reduces to its
plaquette terms. In analogy with the Abelian case
\cite{tagliacozzo_optical_2012}, we exploit the bi-partite nature of the lattice. We imagine coloring
the plaquettes red and black in a checker-board pattern. Next we consider the Hamiltonian (\ref{eq:ham}), but including only half the terms,
{\em e.g.} those on the black plaquettes. With this choice, the model
becomes exactly solvable. As illustrated in Fig. \ref{fig:wc-conf}~a), the ground state $\ket{\Omega}_0$ is a product state of single
plaquette configurations,
\begin{equation}
\ket{\Omega}_0 =\prod_p \ket{\phi}_p, \, \text{with } \ket{\phi_p}=\frac{1}{\sqrt{2}}\left(\ket{\lambda_p}+\ket{\rho_p}\right),\label{eq:Omega}
\end{equation}
where
$\ket{\rho_p} = \ket{1}_{p_1} \ket{0}_{p_2} \ket{0}_{p_3} \ket{1}_{p_4}S_{p_1,p_2} S_{p_3,p_4},$ and
$\ket{\lambda_p} = \ket{0}_{p_1} \ket{1}_{p_2} \ket{1}_{p_3} \ket{0}_{p_4} S_{p_1,p_4} S_{p_2,p_3}$.
We separate the ``position'' part of the Hilbert space from its ``spin''
part by writing the states as elements  of ${\cal P}^{\otimes 4}
\otimes {\cal S}^{\otimes 4}$. Both $\ket{\rho}$ and $\ket{\lambda}$ are represented in Fig. \ref{fig:ingredients}~c).
The state $\ket{\Omega}_0$ factorizes into 
resonating-dimer states. Furthermore, each link, as a consequence of gauge invariance, is entangled with the rest of the system.

We now turn to confinement. Adding a pair of spatially separated  external charges 
rearranges  the singlets into strings connecting the charges.
Each string causes long-range entanglement (LRE) between
the  charges, a distinctive feature of  non-Abelian LGT.
In an Abelian LGT, indeed, a single string does not induce entanglement, since  typically
it involves  flipping a line of spins \cite{levin_string-net_2005}. There, the unique source of  LRE between charges is caused by the linear superposition of several orthogonal string states, present also here.

The ground state is indeed a superposition of  string states along  paths determined by both gauge invariance
and energy minimization. A string passing through a plaquette  increases its energy 
by $\delta E \propto \Delta/g$. Hence, strings touch as few plaquettes as possible. The number of excited
plaquettes is proportional to the inter-charge distance, 
{\em i.e.} the charges are
confined with a string tension proportional to $  \Delta/g$ .
This phenomenon is equivalent to the chromo-electric
flux-tube expected in QCD between two colored charges. 
 The simplest system exhibiting such behavior
consists of only two plaquettes (Fig. \ref{fig:wc-conf}~b).

At \emph{strong-coupling}, the plaquette term in (\ref{eq:ham}) may be neglected. The ground state is the configuration with all the position qubits in the state $\ket{0}$. Hence at any site $s$ the
spin qubits on $s_3, s_4$ form a singlet, see Fig \ref{fig:wc-conf}~c) left. This is a product state
of entangled ``half-plaquettes", of the form $\prod_s \ket{0}_{s_3} \ket{0}_{s_4} S_{s_3, s_4}$.

If we now insert two static charges, a line of singlets must
readjust. As consequence of (\ref{eq:g-inv-ext}), the two spin qubits $Q_{\tilde{s}_3}$, $Q_{\tilde{s}_4}$, originally forming the singlet at $\tilde{s}$, rearrange.
 One qubit, say $Q_{\tilde{s}_4}$ forms a singlet with the external charge, while
the position state associated to the other qubit $Q_{\tilde{s}_3}$ changes from $\ket{0}$ to $\ket{1}$, i.e. the qubit moves to the opposite end of the link.  
There, by (\ref{eq:g-inv}), it is forced to form a singlet with one of the two qubits of the same cross. 
The process repeats until one of the displaced qubits reaches the the second external charge. The result is a gauge-invariant string, stretching between the charges. 
The string is a line of qubits in state $\ket{1}$. The energy of each of them increases by $2 \Delta$; therefore, the static charges are confined. 
Note that the the actual ground state is a superposition of orthogonal strings states inducing  LRE between the external charges. 
However, each single string already entangles the two charges, footprint of a non-Abelian LGT.

\section{Realization through Rydberg atoms}

We implement the $SU(2)$  GM using cold atoms loaded in an optical lattice.
We start by describing  a generic  scheme that works for arbitrary values of $g$,  and then consider  a simplified scheme suited to study
the $g \to \infty$ limit.

\emph{Generic scheme}.
We distinguish two cases. i) The pure gauge theory where atoms only encode the gauge boson degrees of freedom on the links, and 
ii)  gauge fields interacting with matter, where  we need extra atoms to encode the charges at the sites.
%that is we need  at least one extra qubit per site. 
We generically refer to all these atoms as ``ensemble'' atoms.  Since all local Hilbert spaces are  tensor product of qubits, each of them is represented by two (long-lived) hyperfine states of one atom.

The Hamiltonian (\ref{eq:ham}) and symmetry projections operators (\ref{eq:g-inv}) act on at least 8 neighboring atoms. We engineer both sets of operators using the mesoscopic Rydberg gates \cite{Muller09,Weimer10}.  The idea is to add an auxiliary two-level system as a ``control'' atom. The control acts as a switch that turns on and off the interaction between the ensemble atoms, i.e. 
simultaneous Rabi transfer between two hyperfine levels of each of the atoms (SRT). Such operations are induced by laser pulses.
In practice, when the control is initialized in the logical state $\ket{0}_c$ nothing happens, while,
%as all the ensemble atoms are dark states of laser pulses (Electromagnetic induced transparency scheme). 
 once initialized in $\ket{1}_c$, the control is excited to a specific Rydberg state, and causes  SRT on the ensemble atoms   
within its blockade radius. 
The operators in both (\ref{eq:ham}) and (\ref{eq:g-inv}) can be decomposed in linear combination of these SRT (see below and Methods \ref{sect:evol}).
As we only use internal degrees of freedom, the atoms are assumed to be frozen in a Mott state.
Note that the (Zeeman) energy splitting between the logical states (which can be  controlled with a magnetic field) and the lattice depth (controlled by laser intensity) can be taken sufficiently large  to minimize imperfections due to the temperature of atomic sample (see sect. \ref{sect:temp}). In fact,  for ideal gates the simulated temperature of the GM would be zero. 

Thus, the requirements for the implementation of (\ref{eq:g-inv})  and  (\ref{eq:ham}) through Rydberg gate are,  
a) deep optical lattice, loaded with b) two ensemble atoms per link, i.e. the four states of the gauge boson,  and with the appropriate number of matter atoms at each site; 
c) one control atom for each cross (at each site) and plaquette (inside it); 
d) both ensemble and control atoms have two logic (sufficiently split hyperfine) states that can be excited to (different) Rydberg states by laser pulses;
e) the lattice spacing should be tuned such that the ensemble atoms of crosses and plaquettes are physically located inside the blockade radius of their respective control.

\begin{figure}
\includegraphics[width=7cm]{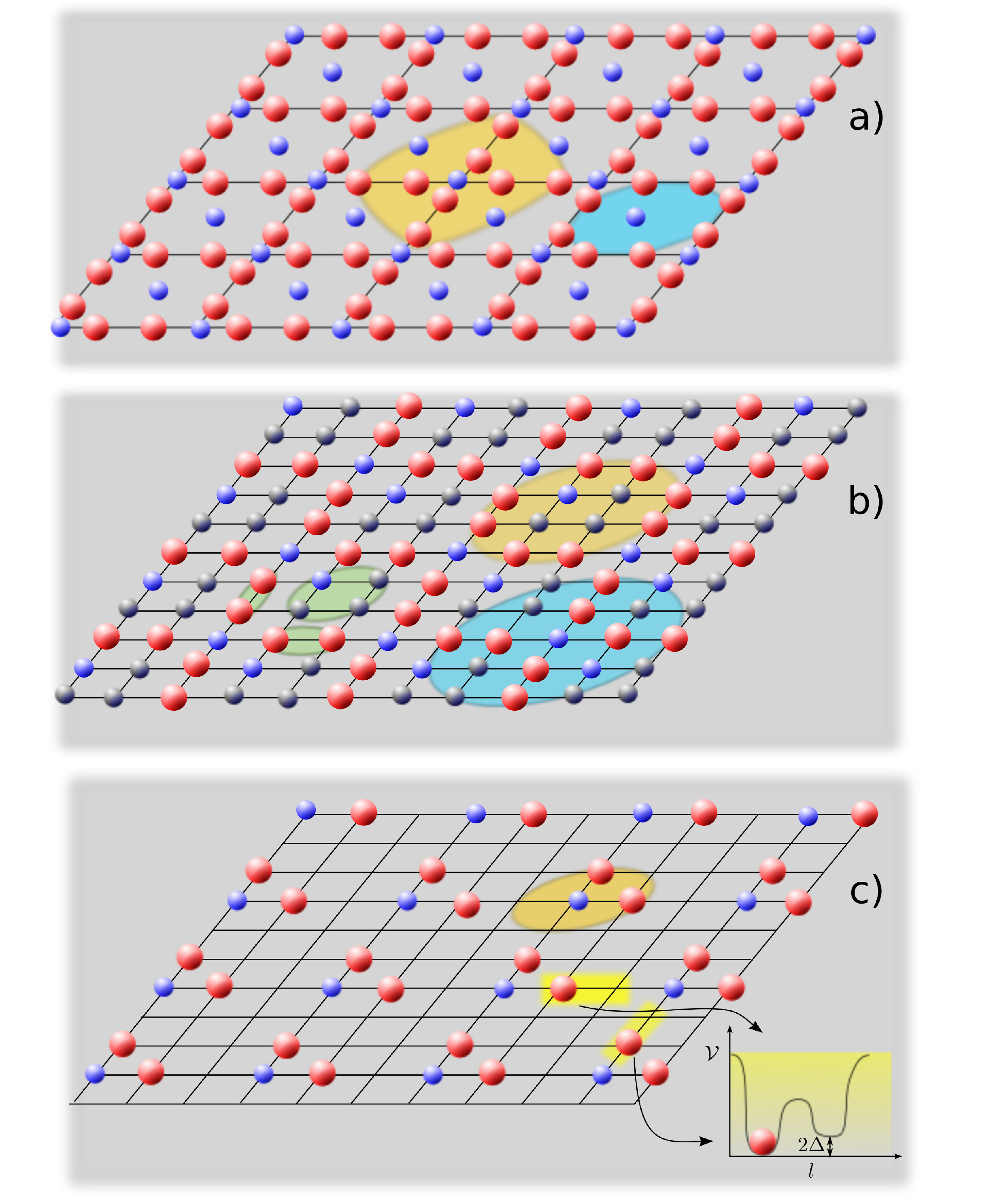}
\caption{\label{fig:opt_lattice}\emph{Lattices of Rydberg atoms}
needed for simulating SU(2)-GM.  {a)}
{\bf Holographic lattice scheme} for the simulation of the pure LGT at
weak coupling. Blue spheres represent the ``control'' two-level
atoms, while red spheres are the ``ensemble'' two-level atoms. The blockade radii for
a plaquette-control atom  and the cross-control atom are shaded in cyan and orange, respectively. {b)} {\bf A different
lattice scheme} has to be used to introduce static charges. It is made from
super-sites including four two-level atoms (shaded  green circle). One of these is a cross-control atom (blue) while the other
three are used to encode static charges (black). The links of the gauge
magnets, which connect these effective sites, have two atoms
each (red spheres on shaded green oval). Atoms used inside the blockade
region are  shaded in cyan and orange. {c)} {\bf Simplified lattice for  strong coupling} simulations.
There is one atoms per link (encoding the spin qubit) and  a double-well
potential at each link (whose wells encod the position qubit); following (\ref{eq:ham})
the relative height of the two potential wells differs by $2\Delta$. Since spin qubits in the ground state are in $\ket{0}$, 
corresponding to the left (lower) well of the potential for $x$ ($y$) links, the right (up) well is empty and the system is at half-filling. 
In order to implement (\ref{eq:g-inv}) the blockade radius of the cross-controls can be limited to the first four wells around a site (shaded in cyan). }
\end{figure}

One way to obtain the desired  optical lattice for both i) and ii) is  sketched in Fig. \ref{fig:opt_lattice}~a)-b), respectively, is to use holographic techniques \cite{bakr_quantum_2009}. In Fig. \ref{fig:opt_lattice}~a), atoms are represented as sphere,  ensemble atoms are red and  control atoms are blue. The required blockade radii are indicated by shaded regions in  cyan and orange. 
%For simulating matter, one also needs to also encode the static charges. A possible lattice configuration used to achieve this is presented in Fig. \ref{fig:opt_lattice}~b).
In presence of matter, a possible lattice configuration is shown in  Fig. \ref{fig:opt_lattice}~b).  
Effective sites (shaded in green) now involve four atoms. Three of them encode the charges (black) while the fourth is a control atom (blue). Links (shaded in green)  connect these effective sites. Again the desired blockade radii are represented by shaded regions, cyan for crosses and yellow for plaquettes. 

% Physically one is free to fix  a) the splitting in energies between the two logical states of the ensemble atoms by using external magnetic fields and b) the lattice depth  by tuning the intensity of the lasers responsible of the optical lattice, so to work in a regime insensitive to the atomic temperature fluctuations see section \ref{sect:temp}.

The experiment we propose aims to detect confinement by measuring the energy of the system as a function of the distance between external charges. A linear growth of the energy  is the footprint of confinement \cite{tagliacozzo_optical_2012}.
We need to prepare  the ground state of the system for any $g$ in  absence and   presence of
static external charges separated by  different distances. We then measure the ground-state energy for each realization, obtaining its dependence on the charges separation.

Before running the experiment for the full 2D GM, we propose to validate the simulator by comparing its outcomes with the known ones for  the exactly solvable scenario described in the previous section, where
the Hamiltonian contains only half of the plaquettes.

The  ground states  are prepared  using  the  adiabatic evolution implemented with the Rydberg gates (see \ref{sect:evol}). 
In order to apply this procedure, we need to  start with a simple state with non-zero overlap with the final ground state. We also need to modify the Hamiltonian by adding  $H_G=-\tilde{\Delta} \sum_{s,\vec{\alpha}=\hat i,\hat j,\hat k}(G_s(\vec{\alpha})+G_s(\vec{\alpha})^{\dagger} )$ that separates of more than  $\tilde{\Delta}$ gauge invariant states from the non invariant ones. At low energies, for $\tilde \Delta \gg \Delta/g$, where $\Delta/g$ is the energy scale of \eqref{eq:ham}, the spectrum is restricted to the physical states.

%%vecchia formulazione
% For example, in the validation step, a good  initial state  can be chosen  as the product state
% $\prod_{l_h}   (\ket + \otimes \ket{1}) \prod_{l_v}(\ket + \otimes \ket{0})$, where $l_h$ and $l_v$ are the horizontal and the vertical links of the lattice, respectively,
% and $\ket +\equiv\frac 1{\sqrt 2}( \ket{0} + \ket{1})$, so that it has  non-zero overlap with $\ket{\Omega_0}$ of (\ref{eq:Omega}).
For example, in the validation step, the product state
$\ket{\Omega_a}\equiv\prod_{l_h}   (\ket + \otimes \ket{1}) \prod_{l_v}(\ket + \otimes \ket{0})$ is a good  initial state  as it has  non-zero overlap with $\ket{\Omega_0}$ of (\ref{eq:Omega}). Here, $l_h$ and $l_v$ are the horizontal and the vertical links of the lattice, respectively,
and $\ket +\equiv\frac 1{\sqrt 2}( \ket{0} + \ket{1})$. 

Note that for the full model \eqref{eq:ham}, we can use $\ket{\Omega_0}$ as the initial state   for the adiabatic evolution, which is now performed by slowly switching on the remaining half plaquettes \cite{tagliacozzo_optical_2012}.  

The evolution is performed for a time ${\cal T}$ ($\propto g/\Delta$) using the Hamiltonian $H(t) =(1- \lambda(t)) H_a + \lambda(t)(H+ H_G)$, 
with the smooth function $\lambda(t)$ that fulfills $\lambda(0)=0, \lambda({\cal T})=1$, and $H_a$ is a gapped Hamiltonian having as a unique ground state the starting configuration.

 %As $[H_G,H]=0$ and $\ket{\Omega_0}$ is the simultaneous minimum of both $H$ and $H_G$,

The adiabatic approximation is justified by choosing ${\cal T}$ large enough compared to the inverse of the gaps of  $H(t)$ which stays finite during the evolution. In practice, as described in detail in the Methods \ref{sect:evol}  the time-evolution operator is applied as a stroboscopic sequence.
We divide the evolution time in a sequence of $N$ short enough intervals $\delta t={\cal T}/N$ such that $\exp i (\delta t H(t)) \simeq \prod_I \exp (i \delta t  \lambda_I(t) H_I) +{\cal O} (\delta t^2)$, where   $H(t)=\sum_I \lambda_I(t) H_I$ and $H_I$ are product of Pauli matrices. Each $\exp (i \delta t  \lambda_I(t) H_I)$  can be implemented by a sequence of SRT on eight qubits (atoms), precisely, with two Rydberg gates and several single-qubit 
rotations (cf. \ref{sect:evol} and \cite{tagliacozzo_optical_2012}). 
Even if Rydberg gates are not perfect and have a finite fidelity, the gap of the Hamiltonian and the adiabatic theorem guarantee that for large enough simulation times the final state is in the desired phase,  see \ref{sect:inf}.

The same procedure is used to obtain the ground state
with two, or more, static charges.
%Also,
% the ground state is an eigenvector of $H$, but not
%with minimal eigenvalue, if we do not limit the Hilbert space to the gauge-invariant subspace.
Their presence is accounted by placing new atoms on the sites of the lattice, for instance in the lattice scheme of Fig. \ref{fig:opt_lattice}~b), and  enforcing (\ref{eq:g-inv-ext}) with a modified $H_G$.

 %Our goal is to demonstrate the confinement of the model and to measure the string tension using the simulator.

In order  to measure the ground-state energy of the system  we use again Rydberg gates.
As discussed  in \ref{sect:evol}, they allow to map the eigenstates of each $H_I$
to states of the control atoms, which  can be then read out by selective 
fluorescence \cite{sherson_single-atom-resolved_2010}  (coupling for example with a laser the state $\ket{0}_c$ to short-lived state).
Sufficient repetitions provide a measurement of   the $H_I$ contribution to $H$. 
By iterating  the same procedure for all the contributions $H_I$ we can sum them  up to $H$ and measure the total  energy of the system.
However, a qualitative detection of  the string of singlets  joining the two
charges, responsible of the confinement, can be done via spin-polarization spectroscopy
\cite{eckert_quantum_2008}.

  What we have described so far is experimentally challenging but can be used to probe confinement {\it in any regime}, even away  from $g \to 0$ where the confinement  is expected from analytical  predictions \cite{orland_lattice_1990}. 
%Experimentally, confinement  would manifest through the shape of the energy-excess around the static charges, as an energy concentration  along a tube joining the two charges. 
%On the contrary, free charges would lead to spreading of the energy, footprint of a dipolar potential. 

In the \emph{strongly-coupled} regime, the decomposition  ${\cal P} \otimes {\cal S}$ of links  makes manifest that the energy depends only
on the position qubit   with  the state $\ket{0}$ ($\ket{1}$) favored (penalized) by $\Delta$. We exploit this to design a
simplified experiment based on  a partial  {\it analog} encoding of links. Each spin qubit is still represented by one atom loaded  in a super-lattice producing  a  double-well potential ${\cal V}$ on each link, see Fig. \ref{fig:opt_lattice}~c). ${\cal P}$ is now encoded by its position in the two wells,  split by an energy $2\Delta$.

When two static charges are added to the system, we need to introduce two extra atoms above the half filled
ground state. The creation of the strings described in the previous section 
can be mimicked by driving the system with $e^{-i\int_0 ^{\cal T}  \lambda(t)H_G  {\rm dt}}$,
$\tilde\Delta>2\Delta$,  while simultaneously inducing in-well atom
oscillations via AC-shaking \cite{ACshaking} of the lattice at
$45^{\circ}$.
Adjusting the intensity of the shaking, we allow for the
adiabatic adjustment of the atoms, which then freeze in minimal
energy configurations compatible with gauge invariance.
 The strings can be observed by direct imaging of the atoms positions, e.g.,
 by joining those atoms found at the right (up) end of the $x$ ($y$) links.
 This also provide the quantitative measurement of the energy needed to assess confinement.

The hybrid encoding of gauge bosons allows to reduce the complexity of the simulation at the level of the digital proposal for simulating Abelian theories  \cite{Weimer10,tagliacozzo_optical_2012}.
Contrary to the generic regime, the simplified setup  implies the existence of an upper bound on the acceptable atomic temperatures $T$,   $k_B T \ll  \Delta < \tilde \Delta$. However, this condition can be satisfied in current state-of-art experiments. (see \ref{sect:strong}).

Summarizing,  we have proposed a quantum simulator of non-Abelian
LGT, based on non-Abelian GM (for similar proposals see \cite{Dalmonte12,Cirac12}). 
We have identified the
mechanism producing charge confinement in this model. We have  characterized its intrinsic non-Abelian nature through the part of the LRE  
generated by a single string configuration. We have designed
experiments that allow to prepare the ground state and mimic the
physics of charge confinement. This is only the first step towards
the full quantum simulation of full-fledged  QCD.
Note that  a slight modification of the model considered here allows for a relativistic dispersion relation as required by QCD \cite{orland_lattice_1990}.
 The physics we
described here is dominated by the presence of singlets, that play
a fundamental role in  high-temperature superconductivity.  We
foresee that the experiments we propose provide also new insights
in this area (see also \cite{trebst_d-wave_2006,nascimbene_experimental_2012}).
An interesting development would be to apply the ideas of \cite{abanin_measuring_2012,cardy_measuring_2011} to measure in the experiments the intrinsic LRE carried by a single chromo-electric string.
%Note that in both cases the number of sequential Rydberg gates to be performed in a time step is 36, i.e., small enough to satisfy $V \gg $ thermal excitations $\sim 1/(100$ Rydberg gate times) for current experiments (see suppl.).

We acknowledge support from TOQATA (FIS2008-00784),
FP7-PEOPLE-2010-IIF ENGAGES 273524, ERC QUAGATUA, and  EU AQUTE.

\section{Methods}

\subsection{Using Rydberg gates}\label{sect:evol}

Mesoscopic Rydberg gates are used both for the ground state preparation and 
its energy measurement.
The ground state for different charge backgrounds $\ket{\Omega}$, is obtained as the ground state of a generalized  Hamiltonian that include a term forcing gauge invariance, $\tilde H=H+H_G$ (see main text).
We start the adiabatic preparation  from an easy-to-prepare unique ground state  $\ket{\Omega_a}$ of an  Hamiltonian $H_a$, such that $\bra{\Omega_a}\ket{\Omega}\neq0$,
 e.g  $H_a= \sum_{l_h} (\Gamma_0-\Gamma_5)+ \sum_{l_v}(\Gamma_0+\Gamma_5)$.
During the evolution, $H_a$ is slowly substituted  by $\tilde H$,  $\tilde H(t)= (1- \lambda(t))  H_a + \lambda (t) \tilde H$ during  total time ${\cal T} \gg 1 /\Delta$. 

In order to implement the time evolution under $\tilde H(t)$, we decompose it in tensor products of Pauli operators.  
For simplicity let us focus just on a single plaquette. It acts on eight atoms encoding eight qubits
\begin{equation}
\tilde H(t) = \sum_{I} \lambda_{I}(t) H_{I},\  H_{I}= \sigma_{i_1}^{1}\otimes \cdots \otimes \sigma_{i_8}^{8}, \label{eq:sigmas}
\end{equation}
$I= (i_1 \cdots i_8)$, and $i=0,1,2,3$, with $\sigma_0=\mathbb{1}$.
For each $H_I$, the plaquette state decomposes as
\begin{equation}
\ket{\psi}= c^I_+\ket{\psi^I_+} +c^I_- \ket{\psi^I_-}, \label{eq:st_dec}
\end{equation}
with $H_I \ket{\psi^I_\pm}  = \pm\ket{\psi^I_\pm}$, and $|c^I_-|^2 + |c^I_+|^2=1$.

The evolution  is approximated  by  a sequence of short steps, of a duration $\delta t ={\cal T}/N$ each 
\begin{multline}
W({\cal T})=\left( e^{-i\int_0^{\delta t} \tilde  H(t) {\rm dt}}\right)^N =\cr
\Bigl(\prod_{I}   e^{-i \lambda_I \int_0^{\delta t}   H_I(t) {\rm dt}}\Bigr)^N + {\cal O}(\epsilon^2)
=\Bigl(\prod_{I} W_I(\delta t)\Bigr)^N + {\cal O}(\epsilon^2),\label{driving}
\end{multline}
where $\epsilon \equiv\max_{I,t} (\epsilon(I,t))$ and $\epsilon(I,t)=\lambda_I(t) \delta  t$.\\
The energy is measured by determining each of the $|c^I_{+}|$  as
\begin{equation}
 \langle H \rangle=\sum_I \lambda_I \langle H_I\rangle=\sum_I \lambda_I \left(2|c^I_+|^2 -1\right) \label{eq:energy}
\end{equation}
 with the $\lambda$'s from \eqref{eq:sigmas}.
 Experimentally, we determine a single $|c^I_+|$  by acting with  
\begin{equation}
\tilde G_I=e^{-i\frac{\Pi}4\sigma_c^2} G_I e^{i\frac{\Pi}4\sigma_c^2}= \mathbb{1}_c\otimes \frac{\mathbb{1}+H_I}2 + \sigma_c^1\otimes \frac{\mathbb{1}-H_I}2. 
\end{equation}
on the state $\ket{0}_c\otimes\ket{\psi}$. This gives  $c^I_+ \ket{0}\otimes\ket{\psi_+} + c^I_-\ket{1}\otimes\ket{\psi_-}$, with $\ket{\psi}$ defined in \eqref{eq:st_dec}.
Up to single-qubit rotations of the ensemble qubits, $G_I=\proj{0}_c\otimes\mathbb{1}+\proj{1}_c\otimes H_I$ is the Mesoscopic Rydberg gate  $G=\proj{0}_c\otimes\mathbb{1}+\proj{1}_c\otimes {\sigma^1}^{\otimes 8}$ responsible of the SRT on all the 8 qubits conditioned on the state of the control \cite{Muller09}. 
Thus, at the price of iterating the process enough times, we can couple $\ket{0}_c$ to a short-lived metastable state and measure $|c^I_+|$ by fluorescence.

A single $W_I(\delta t)$ of \eqref{driving}, acting on $\ket{\psi (t)}$ produces  
\begin{equation}
\ket{\psi (t+\delta t)}^I =c^I_+ e^{-i \epsilon(I,t)}\ket{\psi^I_+} +c^I_-e^{+i \epsilon(I,t)} \ket{\psi^I_-}.
\end{equation}
Experimentally, we realize it by applying  the gate
\begin{equation}
\tilde G_I e^{-i \epsilon(I,t)\sigma_c^3} \tilde G_I=e^{-i\frac{\Pi}4\sigma_c^2} G_I e^{-i \epsilon(I,t)\sigma_c^1} G_I e^{i\frac{\Pi}4\sigma_c^2},
\end{equation}
to  $\ket{0}_c\otimes \ket{\psi(t)}$ \cite{Weimer10}.

The complete experimental sequence is the following: 
a) obtaining the ground state by iterative applications of the $W_I(\delta t)$  and b)   measuring each one of the $H_I$ contributions to the energy. 
The latter inevitably disrupts the ground state,  thus,
 the sequence a) b) needs to be repeated several times. 

\subsection{Effects of the environment and gate errors}

%a)
\subsubsection{The effect of temperature of the atomic sample}\label{sect:temp}

In full digital simulation (for mixed-analog see sect. \ref{sect:strong}), thermal excitations of the atomic sample can be decoupled from GM. Indeed, the logical (hyperfine) states can be Zeeman split by an energy $\delta E \gg K_B T$.
Note that by taking $\delta E\gg \tilde \Delta$, the phase difference due to $\delta E$ is quickly oscillating and cancels from the time-evolution of logic states. 
Furthermore, by working with an optical lattice potential in the Lamb-Dicke regime, we can suppress any excitation due to laser pulses, as lattice modes of the are much more energetic than the recoil energy. The latter 
is dispersed by the lattice itself. Thus, the ``digital'' temperature of GM is zero.   
% 
% 
% The temperature of the atoms is not a real issue, at least for a full digital simulation. Indeed, the logical states, typically hyperfine states of the atom can be split in energy as much as we wish,
% with $\Delta E\gg KT$, thus the temperature of the digital world is zero. Furthermore, we may work in the Lamb-Dicke regime, i.e.  adopt very deep optical lattice potential  such to suppress any undesired motion of the atom.
% In this regime, we expect that the sequence of laser pulses is not producing any excitation, as the modes of the system are much more energetic than the recoil caused by the pulses. 
% In practice, the energy is absorbed by the mirrors via the lasers of the lattice.
% It is worth to notice that the logical states will evolve with a different phase, due to the difference in energy $\Delta E$. However, we assume that $\Delta E$ is greater that any other scale of the Hamiltonian 
% (gauge condition + 
% gauge magnets dynamics) we simulate digitally with Rydberg atoms (or by the mixed approach applied in the strong coupling regime). Thus, the phase shift averages to zero and can be neglected.   
% 
%  In the mixed analog-digital approach proposed to simplify the simulation of the strong coupling regime $g \gg 1$, we have to ask that the temperature of the sample being smaller than the energy scale $\tilde \Delta$ that is 
%  inversionally proportional to the total time  needed to perform a step of the unitary evolution, see sect. \ref{sect:strong}. 
 
%b)
\subsubsection{Effect of spontaneous decay of Rydberg and logical states of atoms}

Regarding  spontaneous decays of the Rydberg, the atoms are in the Rydberg states only when the gates are working. 
As the typical time scale needed to perform the gate is of $\mu s$ and the life time of the Rydberg states can be extended up to 
$ms$ by using sufficiently excited Rydberg states, the spontaneous decay is quite rare and is included in the mechanisms which determine the (in)fidelity of the gate.

To conclude we notice that the coherence time, i.e. the time for which the simulator can work is only limited by  the life time of the metastable logical states. Such time is of order
0.1-1 s, thus in principle $10^5$-$10^6$ Rydberg gates can be safely applied.

\subsubsection{Fidelity of realistic Rydberg gate}\label{sect:inf}

In the main text we have assumed that Rydberg gate to be ideal. The functioning of a realistic Rydberg gate and its fidelity were first discussed in the original paper \cite{Muller09}, 
where the major source of ``infidelity" was re-conducted to the imperfect blockade due to mutual interaction of two or more ensemble  atoms simultaneously excited to Rydberg states.
% Such fidelity decrease with the number of atoms in the ensemble, or better higher detuning and ratio between control and probe pulses are  required to get higher fidelity. 
As argued in \cite{Weimer10}, such imperfection together with possible others can be modeled as 
\begin{align}
 G_I(\phi) &= \proj{0}_c \otimes e^{i\phi Q_I} + \proj{1}_c\otimes H_I\cr
           &=R(\phi) G_I= G_I R(\phi) \label{gphi}, 
\end{align}
where $R(\phi)=\proj{0}_c \otimes e^{i\phi Q_I} + \proj{1}_c\otimes\mathbb{1}$ and $G_I$ was given above.
We may fix the norm of operator $Q_I$ to 1 such as the parameter $\phi$ measures  the``infidelity" of the gate.
Some consideration on the form of $Q_I$. We may expect that the operator decomposes in a systematic error part (which still depends on $H_I$), and a fluctuating random part, which is different each time the gate acts 
(and which we may suppose independent of $H_I$). Hence, we distinguish the $Q_I$ at different time with a prime (for simplicity below we omit the suffix $I$). 
It follows that for a realistic unitary evolution there is non-zero probability, once initialized the control in the state $\ket{0}_c$, of ending in $\ket{1}_c$.   
Thus, to avoid error propagation, we have to force radiative decay of the control 
from $\ket 1$ to $\ket 0$, at the price that the process becomes dissipative. The evolution is 
\begin{equation}
\rho\to C\rho C^\dagger + D\rho D^\dagger,
\end{equation} 
where %following (\ref{uphi})
\begin{align}
C%&=\left(\frac{\mathbb{1} + e^{i\phi Q'}}2 e^{-i\lambda_I H_I}\frac{\mathbb{1} + e^{i\phi Q}}2 + \frac{\mathbb{1} - e^{i\phi Q'}}2 e^{i\lambda_I H_I}\frac{\mathbb{1} - e^{i\phi Q}}2\right)\cr
 &= \frac 12 \left(\cos \lambda (\mathbb{1} +e^{i \phi Q_I'}e^{i \phi Q_I}) - i \sin\lambda(e^{i \phi Q_I'}H_I + H_I e^{i \phi Q_I})\right),\\ 
D%&=\left(\frac{\mathbb{1} + e^{i\phi Q'}}2 e^{i\lambda_I H_I}\frac{\mathbb{1} + e^{i\phi Q}}2 + \frac{\mathbb{1} - e^{i\phi Q'}}2 e^{-i\lambda_I H_I}\frac{\mathbb{1} - e^{i\phi Q}}2\right)\cr
 &= \frac 12 \left(\cos \lambda (\mathbb{1} -e^{i \phi Q_I'}e^{i \phi Q_I})+i \sin\lambda(e^{i \phi Q_I'}H_I - H_I e^{i \phi Q_I})\right).
\end{align} 
Note that the evolution is dissipative already at first order in $\phi$. Indeed,
\begin{equation}
C+D=  e^{-i\lambda H_I} + \phi \sin \lambda  H_I Q_I + O[\phi^2],
\end{equation}
and, by neglecting $\lambda^2$ terms, we have on average
\begin{equation}
<\dot \rho> \propto -i \lambda[H_I + \frac i2\phi [H_I,\bar Q_I], \rho] + \frac 12\lambda\phi\{\{H_I,\bar Q_I\},\rho\},
\end{equation}
where $\bar Q_I$ represents the systematic error, which depends on the $H_I$.  Note that at first order the evolution of the system under the total 
Hamiltonian $\tilde H$ is simple obtained by summing over $I$.

% Hence, we have two effects: the correction to the coherent evolution that goes with ($i\times$) the commutator, which is the Hermitian part of $i \sum_i H_I \bar Q_i$, and a dissipative term which 
% depends on the anticommutator. 
Thus,  
the ``infidelity'' of the gate has two consequences. First, we are not preparing the groundstate of $\tilde H$ but of a perturbed one. 
% The role of the perturbation strongly depends on the properties of the phase we are pointing to. If such state is well separated in energy we may argue using the adiabatic theorem that nothing serious will happen, while, for instance,
% in presence of competing phases very closed in energy the system will be very sensible to $\phi$. 
% Thus, it is not possible to make an assessment a priori on the required fidelity, if the model we are going to simulate is not known.
However, since we are interested in a gapped phase, the adiabatic theorem ensures that the two states are very similar, so that   error is only linear in $\phi$ for $\phi\ll 1$.
Note that $\phi$ depends on the efficiency of the Electromagnetic Induced Transparency employed in the Rydberg gate and can be about $10^{-2}$ \cite{Muller09}.
% Indeed, for a gapped phase the gap is of order one (in $\lambda$ unit), thus is sufficient to require $\phi\ll 1$ ($\phi$ in the Rydberg gate is control by the efficiency of the EIT).
% However, for a gapless in the thermodynamic limit we are limited to the simulation of small lattices (which always have finite gap).
% In fact, a very similar situation appears for classical numerical simulation.
%  
% {\bf some words on validation problem?}
%It is worth to notice that the effect of the $\bar Q_i$ can be mild by the sum over all the terms of the Hamiltonian as they are in principle different.

The second consequence is dissipation. After the adiabatic evolution we have a mixed state. As $H_I$ and $Q$ have finite norm, $|\{H_I,\bar Q_I\}|$ admits one or more states which are eigenvectors with maximum eigenvalue. Thus, each dissipative term tries to drive the system to such states. As the different $\{H_I,\bar Q_I\}$ are expected to be not commuting for different $I$, we may conjecture that their action is to drive the system to thermal state of the deformed Hamiltonian above, with a temperature of order $\phi\Delta$, and, thus, negligible. 

\subsection{Strong coupling experiment} \label{sect:strong}

The strong coupling scheme we propose involves the digital simulation of $H_G=\tilde\Delta\sum_s H_s$ only, $H_s\equiv$ cross operator  at site $s$. 
Due to the analog encoding of the position qubits of the links, each $H_s$ acts only on four atoms  
\begin{eqnarray*}
&&H_s=\sum_{j=1}^3 (e^{i\sigma^j})^{\otimes 4}= 2(\cos \alpha)^4 {\sigma^0}^{\otimes 4} + \cr
& & \ \ \ -2 (\cos \alpha\sin \alpha)^2 ({\sigma^j}^{\otimes 2}\otimes{\sigma^0}^{\otimes 2} + 5 \text{perm.}) + (\sin \alpha)^4 {\sigma^j}^{\otimes 4},
\end{eqnarray*}
thus, we have to engineer only 18 terms with Rydberg gates.

The analog encoding of the position qubit and of its dynamics has also other two consequences. First, thermal fluctuations of the atomic sample are coupled to the GM, thus, contrary to the fully digital scheme above, the absolute value of $\Delta$ and $\tilde \Delta$ does matter. Second, the digital time ${\cal T}$ corresponds to the physical time. 
Thus, we may estimate the highest $\tilde \Delta$ we can engineer by identifying $\delta t$ as the time employed to perform $e^{-i\epsilon H_s}$, $t_{step}$. 
Due to Trotter, $\tilde \Delta =\frac {\epsilon}{t_{step}}$. As each SRT requires two Rydberg gates, $t_{step}= 18\times 2 \times t_R\sim 10^{-5} s$, whether we assume the delay of the Rydberg gate $t_r\sim 1\mu s$.   
It follows that we can neglect the temperature $T$ of the sample for $K_B T\ll \Delta \ll \tilde \Delta \sim 10^{-7}$ $^\circ K$. 

We conclude with few comments on AC-shaking procedure. 
% As explained in the main text, we allow the system to reach the minimal energy configuration during the adiabatic evolution by inducing an hopping between the wells forming the links. In this way, 
% the value of the position qubit, {\emph i.e.} the position of the two-levels atom in each well, can change to the less energetic one while satisfying  the  new gauge condition in presence of charges. The gauge condition determines 
% the configuration of the gauge qubits, forming singlets between pairs  of them.  
The adiabatic modulation of AC-periodic forcing of the lattice provides the hopping the atoms need to adjust to the ground state.
Such hopping is controllable and selective, as it appears only if the frequency of the forcing $\omega$ is commensurable with $2\Delta$ 
\cite{eckardt_ac-induced_2007}. 
This avoids unwanted tunnelings as, for instance, of the control atom to the link wells and vice-versa.
Next-to-nearest neighbor hoppings between different links are strongly suppressed by the exponential decay of Wannier functions.     
In order to induce the same rate of in-well oscillations for horizontal and vertical links, the periodic forcing $\vec F$ is chosen to be parallel to the plane at a $45^\circ$ angle, 
$\vec F \parallel (1,1,0)$.   
%

% \bibliographystyle{apsrev4-1}
% %\nocite{*}
% \bibliography{non-abelian}

\end{document}